\documentclass{emulateapj}

\usepackage{graphicx}
\usepackage{epstopdf}
\usepackage{natbib}
\usepackage{amsmath}
\providecommand{\e}[1]{\ensuremath{\times 10^{#1}}}
\usepackage{url}

\shortauthors{Davenport et al.}
 
\begin{document}

\title{Multi-wavelength characterization of stellar flares on low-mass stars using SDSS and 2MASS time domain surveys}

\author{James R.~A. Davenport\altaffilmark{1,2},
Andrew C. Becker\altaffilmark{2}, Adam F. Kowalski\altaffilmark{2}, Suzanne L. Hawley\altaffilmark{2},  Sarah J. Schmidt\altaffilmark{2},  Eric J. Hilton\altaffilmark{2,3}, Branimir Sesar\altaffilmark{4}, Roc Cutri\altaffilmark{5}}

\altaffiltext{1}{Corresponding author: jrad@astro.washington.edu}
\altaffiltext{2}{Department of Astronomy, University of Washington, Box 351580, Seattle, WA 98195, USA}
\altaffiltext{3}{Current Address:  Institute for Astronomy and Department of Geology and Geophysics, University of Hawaii, 2680 Woodlawn Dr., Honolulu, HI 96822, USA}
\altaffiltext{4}{Division of Physics, Mathematics and Astronomy, Caltech, Pasadena, CA 91125\label{Caltech}} 
\altaffiltext{5}{Infrared Processing and Analysis Center, California Institute of Technology, Pasadena, CA 91125, USA}

\begin{abstract}
We present the first rates of flares from M dwarf stars in both red optical and near infrared (NIR) filters. We have studied $\sim$50,000 M dwarfs from the SDSS Stripe 82 area, and 1,321 M dwarfs from the 2MASS Calibration Scan Point Source Working Database that overlap SDSS imaging fields. We assign photometric spectral types from M0 to M6 using $(r-i)$ and $(i-z)$ colors for every star in our sample. Stripe 82 stars each have 50-100 epochs of data, while 2MASS Calibration stars have $\sim$1900 epochs. From these data we estimate the observed rates and theoretical detection thresholds for flares in eight photometric bands as a function of spectral type. Optical flare rates are found to be in agreement with previous studies, while the frequency per hour of NIR flare detections is found to be more than two orders of magnitude lower.  An excess of small amplitude flux increases in all bands exhibits a power-law distribution, which we interpret as the result of flares below our detection thresholds. In order to investigate the recovery efficiency for flares in each filter, we extend a two-component flare model into the NIR. Quiescent M0--M6 spectral templates were used with the model to predict the photometric response of flares from $u$ to $K_s$. We determine that red optical filters are sensitive to flares with $u$-band amplitudes $\gtrsim$2 mag, and NIR filters to flares with $\Delta u \gtrsim$4.5 mag. Our model predicts that M0 stars have the best color-contrast for $J$-band detections, but M4--M6 stars should show the highest rate of NIR flares with amplitudes of $\Delta J\sim$0.01 mag. Characterizing flare rates and photometric variations at longer wavelengths is important for predicting the signatures of M dwarf variability in next-generation surveys, and we discuss their impact on surveys such as LSST.
\end{abstract}


\section{Introduction}
The study of flares on M dwarfs has a rich history, spanning more than seven decades \citep[e.g.][]{1940ApJ....91..503V, 1949PASP...61..133J}. The similarity of these events to resolved flares on the Sun was recognized many years ago \citep[e.g.][]{lovell1969}. The seminal paper by \cite{kunkel1970} laid the groundwork for our present understanding of the broad multi-wavelength properties of M dwarf flares. The basic two-component flare spectral model (H recombination + warm blackbody continuum) presented in \cite{kunkel1970} is useful for estimating the total energy and stellar surface extent of flares in the optical regime \citep{slhadleo, k10}. However, this model does not specifically address the detailed spectral features observed during flares, such as line emission from H, He I, He II, and Ca II H\&K, nor the evolution of the spectral energy distribution (SED) as the flare cools and decays \citep{2008A&A...487..293F}. A partnership of multi-wavelength photometric and spectroscopic time-domain observing campaigns \citep[e.g.][]{1989SoPh..121...61B,  2003ApJ...597..535H}, and detailed radiative hydrodynamical modeling \citep{2006ApJ...644..484A} are required to extend our understanding of stellar flare emission physics.

\cite{west08} found that the fraction of M dwarfs that show evidence of magnetic activity as defined by H$\alpha$ emission decreases with height above the Galactic plane. This is thought to be an age effect, with the older, less active stars lying further from the plane due to dynamical heating and radial migration within the disk \citep{west08, loebman11}. Flares may play an important role in the main sequence stellar evolution for low-mass stars, and correlating M dwarf flare rates with mass and metallicity promises to yield an effective constraint on the ages of stellar populations.

Models and observations of the initial mass function indicate that M dwarfs are the primary stellar component in the Galaxy by number \citep{2010AJ....139.2679B}. Recent large scale surveys, such as the Sloan Digital Sky Survey \citep[][hereafter SDSS]{york2000}, have photometrically identified millions of M dwarfs \citep{2010AJ....139.2679B}. These stars make excellent tracers of nearby galactic structure due to their high number densities \citep{2007AJ....134.2418B}, and future survey missions, like the Large Synoptic Survey Telescope \cite[LSST;][]{lsst} will detect M dwarfs to nearly the full extent of the Galaxy, using longer wavelength bandpasses and deep repeat imaging. LSST will map the sky approximately 1000 times over 10 years using $ugrizy$ bands, and is expected to detect flares both as variability in previously quiescent stars, and as transient emission from previously undetected sources. This foreground ``fog'' of characteristically blue transient events will prove a significant contaminant to any statistical measurements of variability from cosmological sources \citep{2008ApJ...682.1205R}.

\citet[][hereafter K09]{k09} used the SDSS Stripe 82 time domain photometric data \citep{2007AJ....134.2236S} to look for flares from $\sim$50,000 M dwarf light curves in the $u$ and $g$-bands. From these sparse light curves they were able to recover a flare rate consistent with those found from dedicated photometric monitoring campaigns \citep[e.g.][]{1976ApJS...30...85L}. The rate of flares was found to decrease with peak flare luminosity, and increase with stellar spectral type (redder colors). \cite{hilton10} found serendipitous flares in the time-resolved SDSS spectroscopic data from $\sim$38,000 M dwarfs across the entire SDSS Data Release 6 footprint \citep{dr6}. They confirmed the increase of the flare duty cycle with later spectral type, and further investigated 
the Galactic height dependence of flaring activity, finding that the stars that flared were even more closely confined to the Galactic plane than the stars with H$\alpha$ activity.

The few studies of M dwarf flares in the near infrared (NIR) that have been previously attempted have not found a consistent relationship to the optical emission. \cite{1988ASSL..143..163R}, for example, measured a marginal anti-correlation between optical U-band and infrared K-band photometry of two flares. Similarly, \citet{1995MNRAS.277..423P} found no K-band enhancement during 26 M dwarf flares that were observed simultaneously in the U-band. Observations of flares on the Sun in the NIR offer a more promising avenue. \citet{2006ApJ...641.1210X} studied the luminosity of solar flares at 1.6$\mu$m, the H$^-$ opacity minimum, and found a correlation with hard X-ray and white-light optical emission for two X-class flares. Their data indicated that the NIR band reached peak luminosity after the  X-ray, and was responding to secondary heating, possibly from chromospheric back-warming. 

Modern M dwarf flare studies are expanding both in wavelength and temporal coverage to gain insight into stellar flare physics, as well as the contaminating impact these events have on detecting other sources of variability. \citet{2003ApJ...597..535H} found a correlation between flare energies in the optical, near-UV, and X-ray. \citet{sjs11} have observed flaring M dwarfs with NIR spectroscopy to characterize the correlation between NIR emission lines, such as the H Paschen series, and blue optical broadband flux during flares. Dedicated high time-cadence white-light flare studies from the Kepler mission \citep{2010ApJ...713L..79K} will allow an order-of-magnitude improvement in the study of flare frequency distributions \citep{2011AJ....141...50W} and their possible evolution. \citet{tofflemire2011} have undertaken the most comprehensive simultaneous optical and NIR photometric observing campaign for flares to date to determine the effects of flares on exoplanet detection around M dwarfs.  These high cadence and high precision data place upper limits on the broadband NIR emission from several medium $u$-band flares from M3 and M4 stars with $\sim$3.9 milli-mag photometric precision.

In this paper we seek to extend the analysis of K09 using the time domain data from the 2MASS Calibration Point Source Working Database (hereafter Cal-PSWDB) as well as the SDSS Stripe 82 data. The combination of these data allows us to statistically characterize for the first time the rate of broadband M dwarf flares in red optical and NIR wavelengths. Our data are outlined in \S2. Variability statistics are discussed in \S3. Flare rates as a function of both spectral type and observed passband are presented in \S4. A spectroscopic model of a M dwarf flare is given in \S5, and the predicted sensitivity to finding flares in each band in \S6. Finally a concluding discussion is given in \S7.

\section{Time Domain Databases}
Our data are drawn from the SDSS and 2MASS photometric time domain databases.  Every M dwarf in our study has single-epoch SDSS Data Release 7 \citep[DR7][]{dr7} photometry, used to determine photometric spectral types and distances. Here we describe the selection criteria and spatial matching of our time domain samples.

\begin{figure}[]
\centering
\includegraphics[width=3.5in]{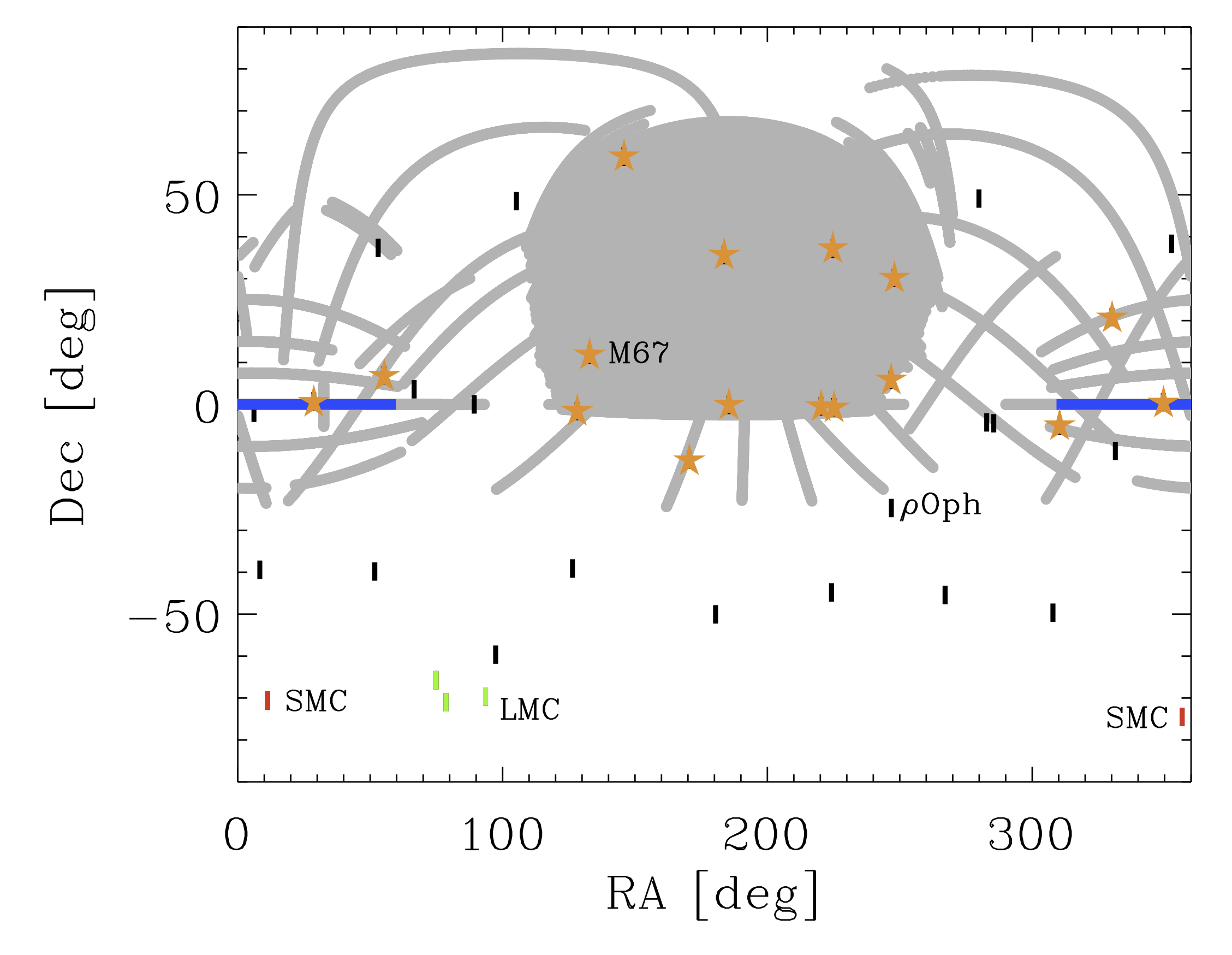}
\caption{The distribution of the Cal-PSWDB tiles across the sky, with a few notable fields labeled. LMC fields are shown in green, SMC in red. Orange stars denote Cal-PSWDB tiles with matches to DR7. The SDSS footprint is shown for reference in grey, with the Stripe 82 data used shown in blue.}
\label{radec}
\end{figure}

\subsection{2MASS Cal-PSWDB}
Our near infrared data come from the 2MASS Cal-PSWDB. Full details are given in the online Explanatory Supplement\footnote{\url{http://www.ipac.caltech.edu/2mass/releases/allsky/doc/seca4\_1.html}} and Cutri et al. (2006). In summary, the repeat imaging was obtained to provide consistent photometric calibration for the 2MASS survey  over the four years of its operation. There were 35 regular calibration tiles spread across the sky, with each tile spanning approximately 8\farcm5 in RA, and 1$^\circ$ in Dec. One of these 35 tiles was visited hourly during survey operations by the 2MASS telescopes, with simultaneous imaging in the $J,H,K_s$ bands. This sequence produced between 562 and 3692 epochs for these calibration fields.  Five additional tiles were imaged in the LMC and SMC with fewer epochs. A total of 191 million point source measurements are contained in the database. A pilot study with the Cal-PSWDB by \cite{becker2008} revealed a 2.6 day low-mass eclipsing binary system. \cite{plavchan08a} used the Cal-PSWDB data to search for periodic variability of low mass stars, finding several eclipsing binary candidates. Additionally, \cite{plavchan08b} found a periodic signature in Cal-PSWDB from a possible triple young stellar object in the $\rho$ Ophiuchus star forming region. 

\begin{figure}[]
\centering
\includegraphics[width=3.2in]{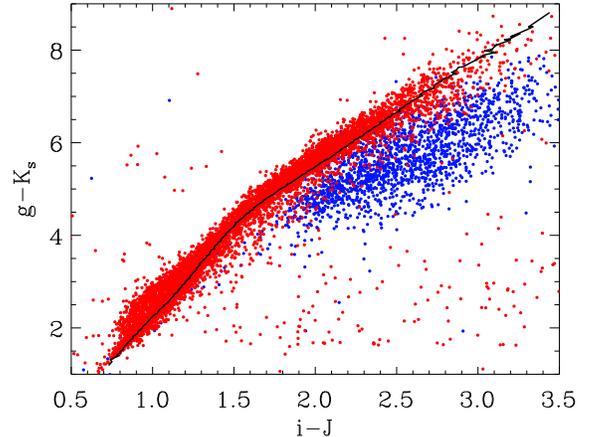}
\caption{A color-color diagram for the 10,000 point source objects with matches in both DR7 and Cal-PSWDB. Stars, as determined by the SDSS OBJ\_TYPE flag, are shown as red circles, galaxies as blue. These two populations nicely separate. The \cite{covey07} fiducial stellar locus is shown for reference, and follows closely the stellar objects in our sample.}
\label{colorcolor}
\end{figure}

We entered the Cal-PSWDB point source measurements into a MySQL database to allow rapid matching of these data to other datasets. The 191 million measurements were clustered into individual objects using the spatial clustering algorithm, \url{OPTICS} \citep{optics}, creating light curves for 113,030 unique objects. Of the 35 primary Cal-PSWDB tiles 16 were coincident with the SDSS Data Release 7 photometric catalog \citep[][hereafter DR7]{dr7}, and we found 11,445 matched point source objects using a 2\farcs5 radius. Figure \ref{radec} shows the  positions of the 40 Cal-PSWDB tiles as compared to the SDSS photometric footprint. An optical--NIR $(g-K_s,i-J)$ color-color diagram for the matched point source objects is shown in Figure \ref{colorcolor}, with 1,963 galaxies and 9,369 stars. The SDSS OBJ\_TYPE flag was used to distinguish between stars and galaxies. We removed the 113 objects with unknown object type (OBJ\_TYPE=0). The distribution of stars in our sample is well fit by the \cite{covey07} fiducial stellar locus.

We imposed magnitude limits of $r<23$ and $z<23$, and standard SDSS photometric flag cuts to ensure good SDSS DR7 photometry. M dwarfs were selected and spectral types assigned using $(r-i)$ and $(i-z)$ colors and the covariance matrix technique as outlined in K09. We found 4,860 stars with photometric spectral types later than M0, with very few stars in the M7-M9 range due to their faint apparent magnitudes. The resulting 4,731 M0--M6 stars had a total of 5.75\e{6} point source measurements in Cal-PSWDB. Only 12 of our stars were in the SDSS spectroscopic M dwarf catalog \citep{westdr7} so we rely solely on the SDSS photometry for classification.

\begin{deluxetable*}{llcr}
\tablecolumns{4}
\tablewidth{0pt}
\tabletypesize{\footnotesize}
\tablecaption{Outline of sample selection cuts for the Cal-PSWDB data.}
\tablehead{
	\colhead{Step}&
	\colhead{Number}&
	\colhead{Cut(s)}&
	\colhead{Description}
	}
\startdata
1. & 11,445 (objects) &  & Cal-PSWDB -- DR7 Match \\
2. & 5.75\e{6} (epochs) 4,731 (stars) & $0\le$SpType $\le 6$, OBJ\_TYPE = 6 & M dwarfs only\\

3. & 2.51\e{6} (epochs) 1,321 (stars) & PH\_QUAL = A, & High S/N Epochs\\
&  & $\sigma_{RA,Dec} < 0\farcs 25$ & Deblending \\
4. & 1.14\e{6} ($JH$ epochs) & $J \le 17$, $H\le16.5$,  $\sigma_J < 0.2$, $\sigma_H < 0.25$& Good Photometry\\
5. & 9.08\e{5} ($HK$ epochs) & $H\le16.5$, $K\le16$, $\sigma_H<0.25$, $\sigma_K < 0.3$  & Good Photometry 
\enddata
\label{sampleselection}
\end{deluxetable*}

We restricted our analysis to epochs with the Cal-PSWDB flag PH\_QUAL = A, which required every $JHK_s$ measurement to have a signal-to-noise ratio of at least 10. To ensure that each DR7 source matched to only a single Cal-PSWDB object, we conservatively removed all objects with a spatial standard deviation in their Cal-PSWDB data of $\sigma_s = \sqrt{\sigma_{RA}^2 + \sigma_{Dec}^2} \ge 0\farcs25$. This cut limited the effects of source confusion and de-blending, but also removed any high proper motion objects from our sample. We additionally excluded epochs where photometric errors exceeded $\sigma_J\ge0.2$, $\sigma_H\ge0.25$, or $\sigma_K\ge0.3$. Our final Cal-PSWDB sample contained 1,321 M dwarfs with $\sim$2\e{6} epochs, for an average of $\sim$1,900 good epochs per star. The selection of our Cal-PSWDB sample is detailed in Table \ref{sampleselection}.

\subsection{SDSS Stripe 82}
In addition to the matched time domain $JHK_s$ data from Cal-PSWDB, we analyzed M dwarfs from the SDSS Stripe 82 repeat scan database \citep{2007AJ....134.2236S} to obtain time domain data in the $ugriz$ bands \citep{1996AJ....111.1748F}. These data were previously used by K09  to search for flares in the $u$- and $g$-bands, and we adopt their sample as described in their Table 2. Briefly, these data contain 50,130 photometrically selected M dwarfs, with 1.9 million total photometric measurements. The stars were required to have good $u$-band magnitudes ($u<22$), which limited the sample to relatively bright M dwarfs. Only 429 of these objects have matches in the Cal-PSWDB database, due to the limited spatial overlap. We measure the flare rate independently in each filter to build statistics using the entire population from each time-domain survey.

\begin{figure*}[t]
\centering
\includegraphics[width=6in]{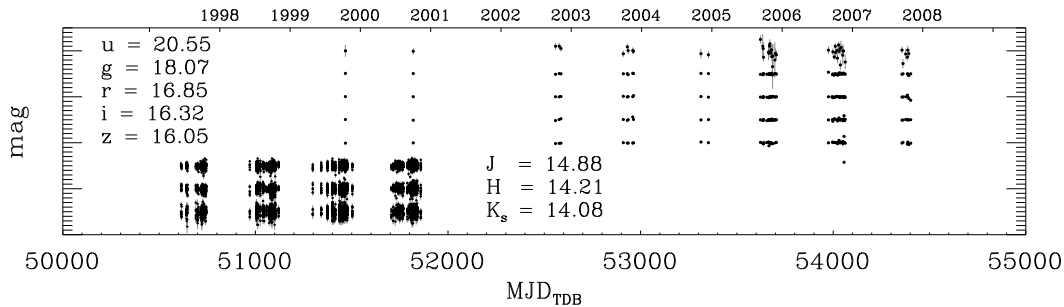} 
\caption{An eight-band light curve for SDSS J231820.76+010141.3, which had both Stripe 82 and Cal-PSWDB time-domain photometry. This object had a photometric spectral type of M0 assigned. Time has been converted for both surveys into BJD TDB. The median good magnitude for each filter is given  beside each light curve. For reference the calendar year is given on the top-axis.}
\label{lc}
\end{figure*}

\begin{figure}[b]
\centering
\includegraphics[width=3.5in]{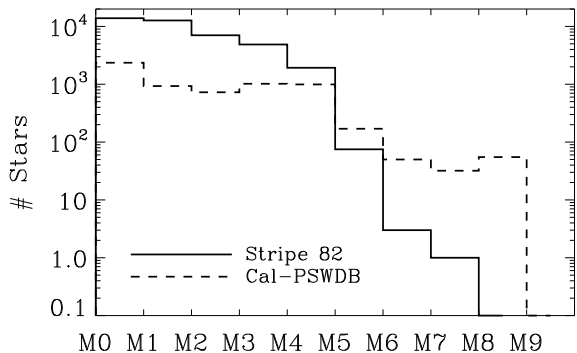}
\caption{The distribution of spectral types for our Stripe 82 and Cal-PSWDB samples. Due to the lack of stars in the M7--M9 bins we have only performed our analysis on M0 through M6 sub types.}
\label{sptype}
\end{figure}

The spatial location of the SDSS equatorial Stripe 82 data is shown in Figure \ref{radec} in blue. Figure \ref{lc} gives an example of the eight-band light curves we have produced for a representative M0 star with both Stripe 82 and Cal-PSWDB photometry, illustrating the differences in time coverage and cadence between the two surveys. This star had 2,922 epochs of data in each of the $JHK_s$ bands, and 70 epochs in each of $ugriz$. The time units for all of our SDSS and 2MASS data have been converted to barycentric dynamical time ($\rm BJD_{TDB}$), and are presented as a modified barycentric Julian date (BJD-2400000.5). Figure \ref{sptype} shows the numbers of stars in each spectral type bin for our Stripe 82 and Cal-PSWDB samples.

\section{The $\Phi$ Statistic}

Discerning true variability from background noise requires sophisticated statistical techniques. Following K09, we utilize a modified Welch--Stetson variability index, $\Phi$, to search for flares in our sample \citep{welchstetson, 1996PASP..108..851S}. This statistic uses measurements in two bands to separate real stellar variability from the intrinsic scatter in a light curve. For the $n$-th measurement of a light curve in $J$- and $H$-bands, for example, $\Phi$ is defined as:
\begin{equation}
\Phi_{JH}(n) = \left[\frac{m_J(n) - \langle m_J\rangle}{\sigma_J(n)} \right]\left[ \frac{m_H(n) - \langle m_H\rangle}{\sigma_H(n)}\right]\,,
\end{equation}
where $m(n)$ is the apparent magnitude at epoch $n$, $\langle m\rangle$ the median magnitude over all epochs, and $\sigma(n)$ the photometric error at epoch $n$.  When the magnitude increases or decreases for {\it both} bands simultaneously, $\Phi$ yields a positive value. Conversely, when the magnitude from one band is higher than normal while the other is diminished, $\Phi$ is negative. For purely random noise between the two filters, this index would yield a symmetrical distribution about $\Phi=0$. Excesses of positive $\Phi$ values represent correlated variability, from events such as flares or eclipses. 

\begin{figure*}[t]
\centering
\includegraphics[width=6in]{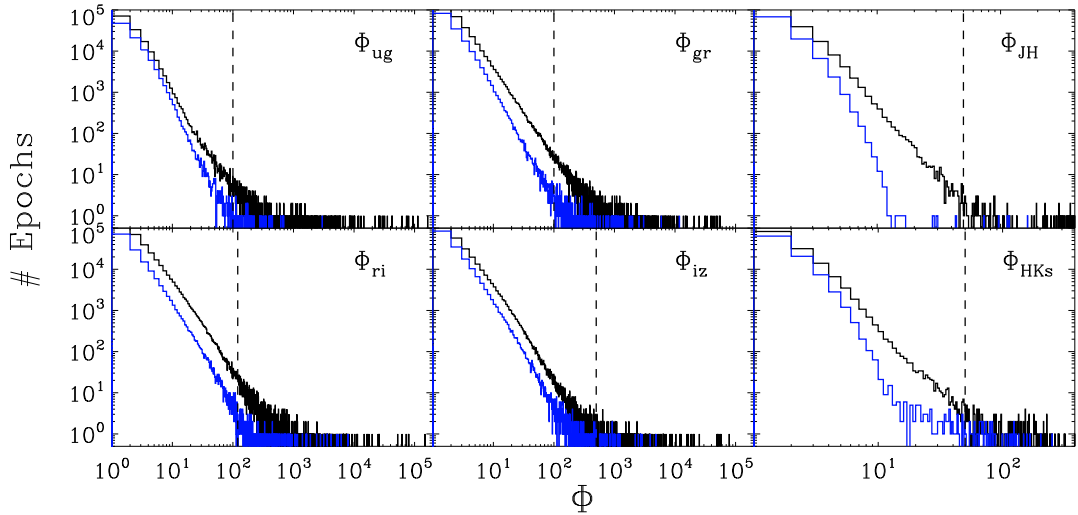}
\caption{Histograms of the two-band variability index, $\Phi$, for each filter combination. The $\Phi^+$, where the flux in both filters is greater than their respective median values, are shown in black. The $\Phi^-$ null distribution has been flipped about $\Phi=0$ for comparison. We subtracted the null distribution from the positive distribution to measure the residual numbers of epochs with  correlated increases in flux between bands. The $\Phi$ thresholds used in each distribution to define flare candidates are shown as vertical dashed lines. These are described in the text.}
\label{phi}
\end{figure*}

We computed the $\Phi$ statistic for all pairs of bandpasses ``adjacent'' in wavelength space  (e.g. $\Phi_{ug}$, $\Phi_{JH})$ for every epoch in our samples. Since the SDSS and 2MASS data were not obtained contemporaneously (see Figure \ref{lc}) we do not compute $\Phi_{zJ}$. Figure \ref{phi} shows the $\Phi$ histograms for all objects in our Stripe 82 and Cal-PSWDB samples. The black line indicates the $\Phi>0$ data, hereafter denoted $\Phi^+$, while the blue line shows the $|\Phi<0|$ data, hereafter $\Phi^-$.

As the sampling cadence of Stripe 82 and Cal-PSWDB surveys was much longer than typical flare timescales of minutes to hours \citep{1974ApJS...29....1M}, our flare search relies on single-epoch outliers in the light curves. In order to act as an effective variability discriminant the two bandpasses making up the $\Phi$ statistic must be imaged nearly simultaneously to capture the same physical event. The 2MASS $JHK_s$ bands were imaged simultaneously \citep{2mass}, allowing the photometry to be directly compared for each 7.8 second exposure. The SDSS used a drift-scanning technique that produced separate 52.4 second exposures in each filter taken in the order $r,i,u,z,g$ \citep{york2000}. As a result the $\Phi_{ri}$ values, for example, come from measurements which were separated by only $\sim$1 minute, while the $\Phi_{gr}$ values are derived from data which were separated by almost 5 minutes. We searched for flares in each of the six $\Phi$ distributions separately, but recognize the limitation that flares with timescales of only a few minutes may not have been imaged by all SDSS bands. As such, these rapid events would not yield large $\Phi$ values, and may fall below our detection thresholds.

To remove large $\Phi$ values from the sample that resulted from a {\it decrease} in flux in both bands, for example due to photon loss from clouds or eclipses, we required all epochs in our investigation to have $m < \langle m \rangle$ for the bluer (shorter wavelength) filter in each of the $\Phi^+$ and $\Phi^-$ distributions. This removed approximately half the epochs in each distribution, as expected. As discussed in K09, the $\Phi^-$ distribution, also referred to as the ``null'' distribution, is representative of uncorrelated noise in the light curve. The $\Phi^-$ distribution was still well modeled by gaussian noise in all filters after the removal of half the epochs. Comparison of the $\Phi^+$ distribution with the $\Phi^-$ yields excesses in the $\Phi^+$ due to candidate flaring epochs.

\section{Optical and NIR Flare Rates}
In this section we explore the selection of flares from our light curves, and the luminosities of the flare candidate epochs.

\subsection{Observed Rates}
As discussed above, given our sparse sampling we assume that each flare is observed in only one epoch of our data. A threshold value for $\Phi^+$ must be chosen from the null distributions to separate flare epochs from the quiescent epochs with statistical fluctuations. If too high a threshold is chosen, we may lose valuable weak signal in the redder bandpasses by disregarding real flare events. However, too low a threshold will add significant false positives to the resulting flare rates. Using the False Discovery Rate method \citep{2001AJ....122.3492M} on the $\Phi_{ug}^-$ distribution, K09 determined that a conservative cutoff for flares was $\Phi_{ug}^+ \ge 100$. This ensured less than 10\% contamination from non-flaring epochs would be included in their final analysis. We performed the False Discovery Rate analysis to limit the contamination to 10\% in all six of our $\Phi^-$ distributions, and determined $\Phi$ thresholds of (100, 100, 120, 500, 50, 50) for $(\Phi_{ug}, \Phi_{gr}, \Phi_{ri}, \Phi_{iz}, \Phi_{JH}, \Phi_{HK_s})$ respectively. The $\Phi^-$ distributions using 2MASS filters in Figure \ref{phi} decline rapidly at $\Phi\sim20$, nearly an order of magnitude lower values than for the optical filters. This is indicative of the lower intrinsic stellar variability seen at these wavelengths, and the larger photometric errors in the 2MASS survey. The $\Phi$ thresholds are shown as dashed vertical lines in Figure \ref{phi}.

\begin{figure}[]
\centering
\includegraphics[width=3.5in]{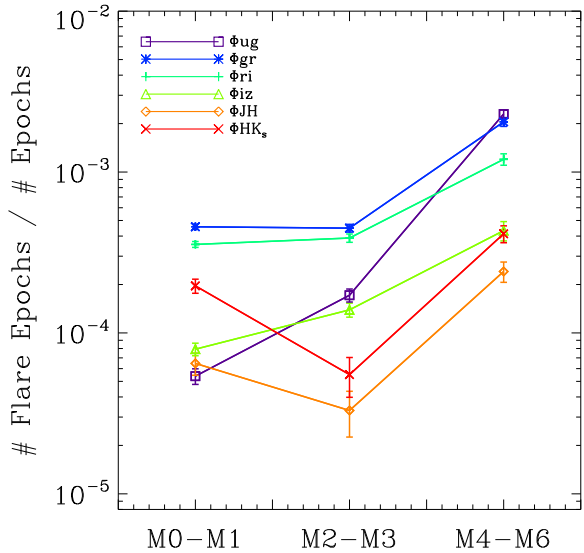}
\caption{The fraction of flaring epochs, defined as the number of good epochs above the $\Phi$ threshold divided by the total number of good epochs, for each filter combination and spectral type bin.}
\label{ffrac}
\end{figure}

The number of flaring epochs for each filter combination was very small compared to the total number of observations. Previous studies such as K09 have shown that flare rates for M dwarfs increase as a function of spectral type. To examine this trend we grouped our sample into three spectral type bins (M0-M1, M2-M3, M4-M6). The ratio of flare candidate epochs to the total number of epochs for each bin is shown in Figure \ref{ffrac}. As expected, all the optical SDSS filter combinations  show an increasing numbers of flares with later spectral type (K09). There were considerably fewer flare epochs detected in the NIR passbands, although the $\Phi_{JH}$ and $\Phi_{HK}$ flare rates fall within the range of the SDSS data in Figure \ref{ffrac}. This is likely due to the weak signatures of flares in these NIR bands, as discussed in \S5. Table \ref{flaretable} presents the total numbers of flares in each band for the three spectral type bins.

\begin{deluxetable}{lcccc}
\tablecolumns{4}
\tablewidth{0pt}
\tabletypesize{\footnotesize}
\tablecaption{Summary of flare candidates in each spectral type bin per wavelength pair.}
\tablehead{
	\colhead{$\Phi$}&
	\colhead{Spectral}&
	\colhead{Candidate}&
	\colhead{Number}&
	\colhead{Flare} \\
	\colhead{}&
	\colhead{Type}&
	\colhead{Flares}&
	\colhead{Epochs}&
	\colhead{Fraction}
	}
\startdata
ug & M0-M1 & 81 & 1.5\e6 & 5.4\e{-5}\\
ug & M2-M3 & 116 & 6.8\e5 & 1.7\e{-4}\\
ug & M4-M6 & 262 & 1.1\e5 & 2.3\e{-3}\\
gr & M0-M1 & 722 & 1.6\e6 & 4.6\e{-4}\\
gr & M2-M3 & 322 & 7.2\e5 & 4.5\e{-4}\\
gr & M4-M6 & 251 & 1.2\e5 & 2.1\e{-3}\\
ri & M0-M1 & 567 & 1.6\e6 & 3.6\e{-4}\\
ri & M2-M3 & 283 & 7.3\e5 & 3.9\e{-4}\\
ri & M4-M6 & 149 & 1.2\e5 & 1.2\e{-3}\\
iz & M0-M1 & 125 & 1.6\e6 & 8.0\e{-5}\\
iz & M2-M3 & 100 & 7.2\e5 & 1.4\e{-4}\\
iz & M4-M6 & 53 & 1.2\e5 & 4.3\e{-4}\\
JH & M0-M1 & 41 & 6.3\e5 & 6.5\e{-5}\\
JH & M2-M3 & 10 & 3.0\e5 & 3.2\e{-5}\\
JH & M4-M6 & 48 & 2.0\e5 & 2.4\e{-4}\\
HK & M0-M1 & 99 & 5.1\e5 & 1.9\e{-4}\\
HK & M2-M3 & 13 & 2.4\e5 & 5.5\e{-5}\\
HK & M4-M6 & 69 & 1.7\e5 & 4.1\e{-4}
\enddata
\label{flaretable}
\end{deluxetable}

\subsection{Flare Luminosities}
To compare our  results in many photometric bands to those of previous studies, we must transform the flares from apparent magnitude enhancements to luminosities. The apparent magnitudes for every epoch and filter in our time-domain Stripe 82 and Cal-PSWDB samples were first converted to observed flux, in units of erg s$^{-1}$ cm$^{-2} \mu$m$^{-1}$, using the calibrations from \cite{2007AJ....134..973I} and \cite{2003AJ....126.1090C} respectively. To estimate the  quiescent luminosity for each epoch, we obtained distance estimates for our stars from the $(\mu_r,r-z)$ photometric parallax relation of \cite{2010AJ....139.2679B}, using the $(r-z)$ color from DR7. Errors in the computed luminosities are dominated by the distance uncertainties.

\begin{figure}[]
\centering
\includegraphics[width=3.5in]{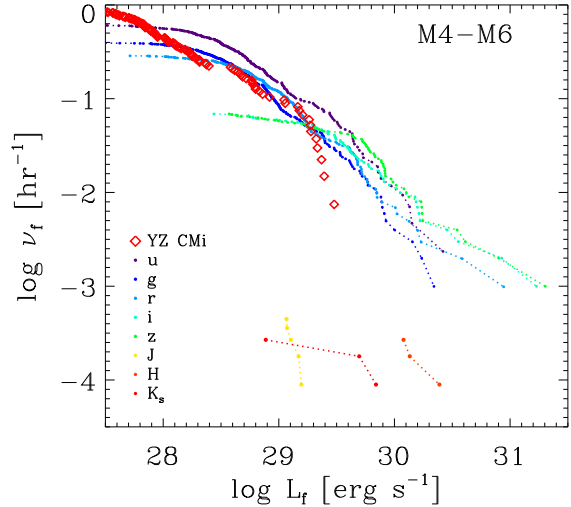} 
\caption{The cumulative flare frequency distributions as a function of observed luminosity in each passband for a sample spectral type bin. Errors in luminosity are dominated by photometric errors and assumptions in the distance. Incompleteness in recovering flares at the low-luminosity end results in the flattening seen clearly in each of the SDSS bands.}
\label{ffd}
\end{figure}

Many previous studies have shown that the frequency of flares decreases as a function of increasing flare energy \cite[e.g.][]{1976ApJS...30...85L} or flare luminosity (K09). For each bandpass we selected the epochs identified as flare candidates in Table \ref{sampleselection}, and found a cumulative distribution in flare luminosity. The frequency per hour that these flares are observed was calculated by dividing the cumulative number distribution by the total observing time in hours for each filter, giving the flare frequency distributions plotted in Figure \ref{ffd}. These show for each filter, $f$, the number of flares per hour of observation, $\nu_f$, as a function of observed luminosity, $L_f$.  The YZ CMi flare frequency distribution from \cite{1976ApJS...30...85L} is shown, and corresponds well with our optical distributions.

The optical ($ugriz$) distributions are very similar, indicating that flares produce smooth continuous variations in these bandpasses. As with Figure \ref{ffrac}, the $JHK_s$-band flare frequency distributions in Figure \ref{ffd} are highly under-sampled. NIR flares were recovered more than two orders of magnitude less frequently than for SDSS optical colors for the M4--M6 bin.
 
\begin{figure}[t]
\centering
\includegraphics[width=3.5in]{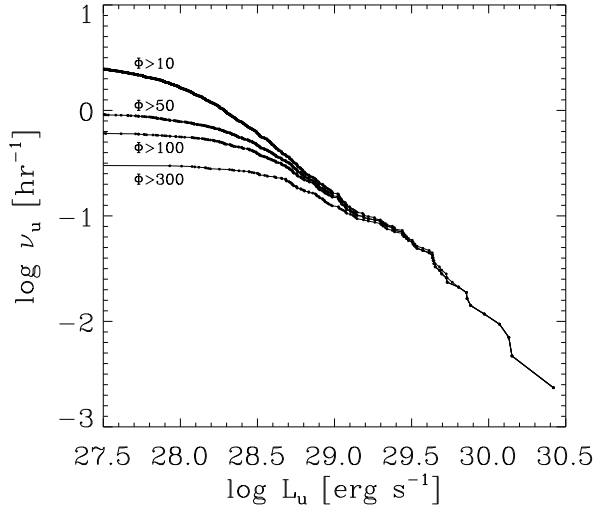}
\caption{The cumulative flare frequency distribution, as in Figure \ref{ffd}, for the $u$-band using four different $\Phi$ thresholds to select flare candidate epochs. Lower thresholds allow more contaminating epochs into our flare distribution, but the power-law slope seen at luminosities above $\sim10^{29}$ erg s$^{-1}$ remains the same.}
\label{phicut}
\end{figure}

The turnover at low luminosities seen in the SDSS filters is due to incompleteness in the low-luminosity flare census. To explore whether the observed power-law distributions (from log $L_f$=29 to 31 erg s$^{-1}$) extend to even lower luminosity flares than our $\Phi$ thresholds recover, we moved the $\Phi^+$ detection thresholds to higher and lower values in the $u$-band. In Figure \ref{phicut} the power-law slope appears to extend to much lower luminosities as a result of the lower $\Phi$ thresholds. We recovered many additional flare candidate epochs, but with a much higher rate of false positives. The high-luminosity end of Figure \ref{phicut} (above log $L_u\sim$29.1 erg s$^{-1}$) is insensitive to the choice of $\Phi$ threshold. We present our flare frequency distributions as observed, but note that it is possible to correct the frequencies for the efficiency in recovering flares as a function of luminosity (E. J. Hilton 2011 PhD Thesis).

\subsection{Residual $\Phi$ Distributions}
\begin{figure}[]
\centering
\includegraphics[width=3.5in]{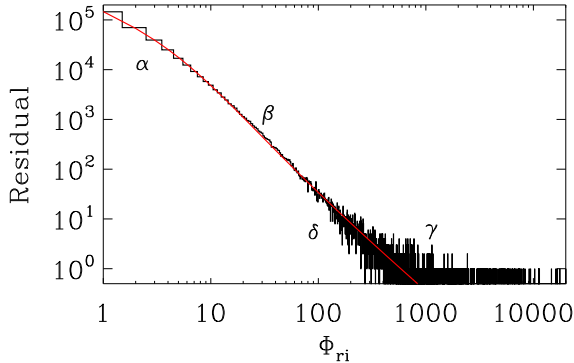}
\caption{The $\Phi_{ri}$ residual (black), and the corresponding fit (red) defined in Eqn \ref{phieqn}.}
\label{resid_ri}
\end{figure}

\begin{figure}[]
\centering
\includegraphics[width=3.5in]{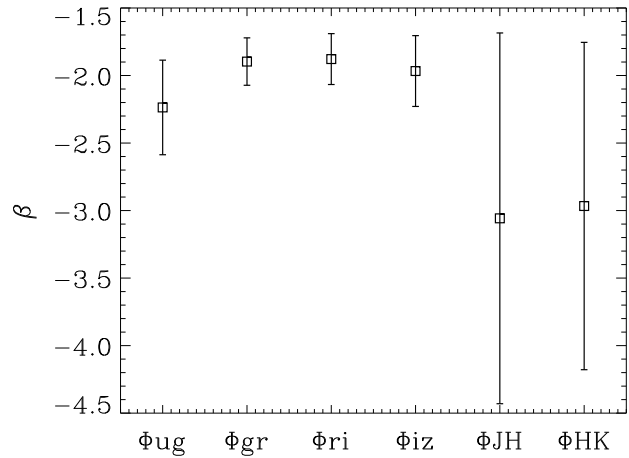}
\caption{The slope of the residual $\Phi$ distributions, as defined in Eqn \ref{phieqn}, for each filter combination. The steeper slope at increasing wavelength is due to the lower intrinsic rate of variability within the photometric errors.}
\label{resid}
\end{figure}

The $\Phi$ distributions in Figure \ref{phi} indicate that there are many more epochs in the $\Phi^+$ distribution than in the $\Phi^-$, showing that the two bandpasses which make up each $\Phi$ distribution preferentially increased in flux together. The excess between the $\Phi^+$ and $\Phi^-$ is also evident at values of $\Phi$ well below the flare candidate thresholds. By subtracting the $\Phi^+$ from the  $\Phi^-$ distributions for each of the six filter combinations in Figure \ref{phi} we are left with residual $\Phi^+$ distributions, as shown in Figure \ref{resid_ri} for $\Phi_{ri}$. These residual distributions follow a roughly power-law profile below the $\Phi^+$ thresholds, with a turn-over at low $\Phi^+$ amplitudes. We fit the residual distributions using the equation
\begin{equation}
\log_{10}Y = \alpha + (\beta\log_{10}X) (10^{-\gamma / (X + \delta)}) \;,
\label{phieqn}
\end{equation}
\noindent
where $Y$ is the $\Phi^+-\Phi^-$ residual, $\alpha$ is the peak amplitude for the residual, $\beta$ a power-law slope at larger luminosities, $\delta$ is the turnover $\Phi$ between the power-law and the flattening at low $\Phi$, $\gamma$ the exponential plateau slope at low $\Phi$, and $X$ is the $\Phi^+$ value. Figure \ref{resid_ri} also illustrates where in the residual distribution each of the fit coefficients constrains. This functional form was chosen to simultaneously fit the residual power-law tail at large $\Phi$, and the turnover at low $\Phi$ values due to random noise dominating. The power-law coefficient $\beta$ is plotted for each of the distributions in Fig \ref{resid}. The $\beta$ values ranged between -1.9 and -3.1. The higher slope in $\beta$ in the $\Phi_{gr}$ fit may be due to the better photometric precision in the $g$ and $r$ bands, compared to the $u$ band, allowing those filters to be more sensitive to flares below the $\Phi$ threshold. The Cal-PSWDB $\Phi_{HKs}$ conversely shows the smallest observed residual, again suggesting that these longer wavelength filters are less affected by intrinsic stellar variability. However, all the fits are consistent with a value of $\beta\approx-2$ within the errors.

The physical cause of this power-law slope, $\beta$, at $\Phi$ values less than our flare cutoff threshold, may be the extension of the increasing flare frequency at lower luminosities. \cite{1976ApJS...30...85L} showed that the flare frequency distribution has a power-law shape, with a turnover at lower energies due to confusion with the quiescent variability, and a fall-off at higher energies due to incompleteness. Their flare detection extended to lower energy flares than our analysis was sensitive to. This trend has been verified for M dwarf flares more recently by K09 and \cite{hilton10}.

There is not a direct transformation to compare between our observed $\Phi$ residual distributions and the flare frequency distributions calculated by \cite{1976ApJS...30...85L} and others. Our residual $\Phi$ distributions contain all the M sub-types combined, for stars across a large range of distances. \cite{2010arXiv1012.0577H} have shown that the flare frequency distribution slopes are similar, though not constant, for different spectral type bins, and between active and inactive stars

\cite{1995ApJ...451..795R} and \cite{1999ApJ...516..916R} found that ``microflares,'' very frequent low-energy rapid flares, were detectable using HST for two very active M dwarfs. By comparing the distribution of fluxes about the median quiescent value in their light curves, the microflares, which are of too low amplitude to be detected individually, could be detected statistically. They found that these microflare events followed a power-law distribution which was steeper than that of the larger resolved flares previously detected. Thus there are even greater numbers of microflares than would be predicted by simply extending the power-law flare frequency to lower energies. 

Our method of finding the residual $\Phi$ distribution yields a similar result, by determining an asymmetry in the light curves via the $\Phi$ statistic due to excess flux enhancements. However we caution that we are not detecting the extremely low-amplitude microflares seen in the HST observations from \cite{1995ApJ...451..795R}. Instead, the residual power in the $\Phi$ distributions are likely a result of the underlying cacophony of low amplitude flares that are below our single-epoch detection limits in each bandpass, but which would be visible in high-precision continuous monitoring programs. It is especially notable that the significant asymmetric distributions were seen in all wavelengths studied ($u$--$K_s$), although we are sampling very different contrasts with the underlying quiescent stellar spectrum, which supports our interpretation of these events as flares. While the flare frequency distributions in Figure \ref{ffd} are under sampled for the NIR filters, the residual epochs with excess flux seen in the $\Phi_{JH}$ and $\Phi_{HKs}$ in Figure \ref{phi}, and their power-law slopes in Figure \ref{resid}, provide a statistically significant detection of small-amplitude flux enhancements. 

\section{Optical and NIR Flare Model}
We have employed a semi-analytical two-component flare model, as first advocated by \cite{kunkel1970}, to explore the completeness of our search for flares in every filter and to compare the observed flare luminosities between filters. We extend the \citet{k10} model which was used in blue optical wavelengths to red optical and NIR bandpasses.

\begin{figure}[]
\centering
\includegraphics[width=3.5in]{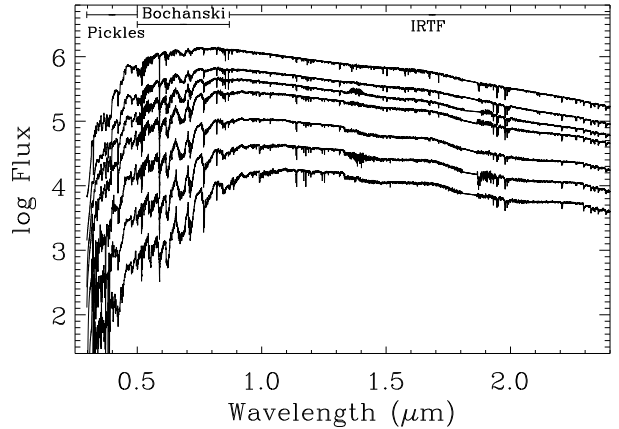}
\caption{The combined 2900\AA\, to 2.5$\mu$m spectral energy distributions for our M0-M6 templates. The ranges in wavelength over which each of the template sources contributes are shown at top.}
\label{sed}
\end{figure}

In order to simulate flux enhancements due to flares in each filter using the two-component model, we built quiescent template M dwarf spectra in the optical and NIR. We utilized flux calibrated spectra from the IRTF stellar catalog \citep{2005ApJ...623.1115C, 2009ApJS..185..289R} for stars with spectral type M0 through M6. The M0 through M6 SDSS templates from \cite{2007AJ....133..531B} were used for the most of the optical regime. The IRTF and Bochanski spectra were normalized and joined at 8500\AA. Since the Bochanski optical templates only go as short as 3825\AA, to fully cover the M dwarf spectra in the $u$-band we used templates from \cite{1998PASP..110..863P}. As these are lower resolution than the Bochanski templates, we used them only from 2900\AA\ to 5000\AA. The Pickles and Bochanski templates are both published in normalized flux units, while the flux-calibrated IRTF data we used in the NIR are in observed flux. We found distances to each of the IRTF M dwarf standard stars from \citet{1952QB813.J45......}, \citet{1995gcts.book.....V}, \citet{2007A&A...474..653V}, and \cite{2009ApJ...704..975J} using the SIMBAD database, and standard radii from \cite{nlds} to convert the observed flux into surface luminosity for each star in the NIR. The normalized optical templates were then scaled to match at 8500\AA, and or resultant optical/NIR spectral energy distributions for the M0 through M6 sub types are shown in Figure \ref{sed}.

\begin{figure*}[]
\centering
\includegraphics[width=6in]{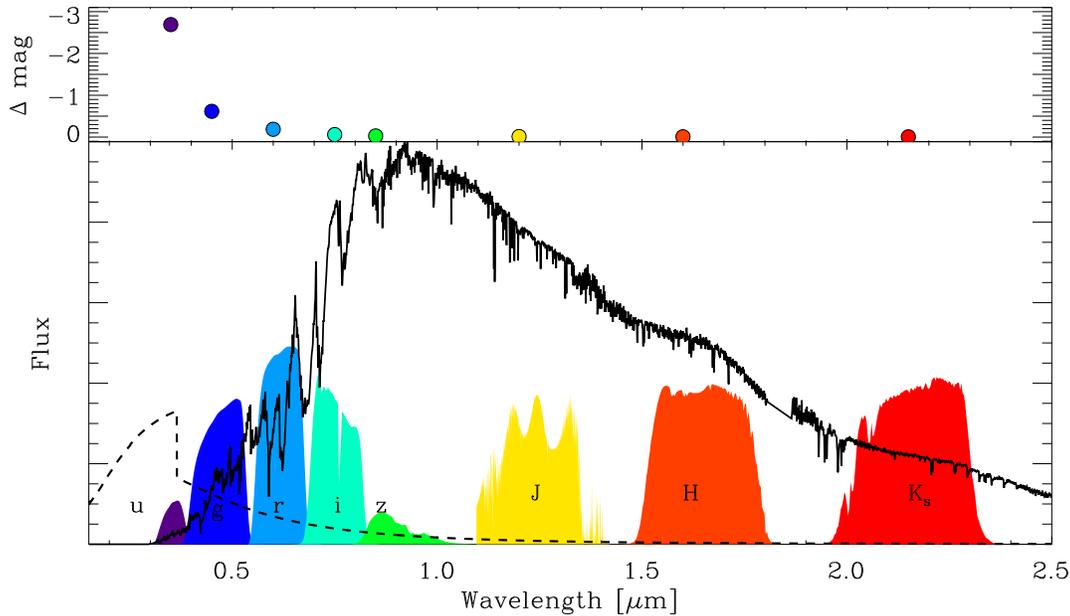}
\caption{Bottom: Our two-component flare model (dashed line) shown against the quiescent M3 optical-NIR stellar SED (black line). The $ugrizJHK_s$ filter passbands are shown for reference. Top: The resulting change in magnitude for each filter from a sample flare with 1\% surface coverage on an M3.}
\label{model}
\end{figure*}

The lower panel of Figure \ref{model} presents our M3 spectrum in arbitrary flux units in black, with the  $ugrizJHK_s$ transmission curves from \cite{1996AJ....111.1748F} and \cite{2003AJ....126.1090C} shown for reference. The overall flare SED shape was defined by a 10,000K blackbody continuum. Previous studies have found temperatures as low as $\sim$8500K \citep{1992ApJS...78..565H} for flares, while \cite{k10} found evidence for temperatures as high as 13,000K. \cite{2008A&A...487..293F} fit a blackbody temperature of 20,000 K during the brief impulsive phase of an enormous flare on CN Leo. These temperatures rapidly declined, however, with values around 7,000--10,000 K for the remainder of the flare event.

Shortward of 3646\AA, we added a second component of H Balmer continuum line emission, which was computed from the \cite{2006ApJ...644..484A} radiative hydrodynamic M dwarf atmosphere model. The Balmer continuum was scaled to have 10 times larger surface area coverage than the blackbody component. The 10:1 scaling was approximately an average of those used by previous studies, with \cite{k10} using values from 3:1 to 16:1. Our idealized model does not include many detailed emission mechanisms that would alter these fluxes, such as line emission from other elements, or other levels of hydrogen. Using measured spectral line emission from M dwarf flares in the NIR  from \citet{sjs11}, we estimate that line emission would contribute an order of magnitude less than our predicted blackbody continuum emission in $JHK_s$ bands, and is therefore negligible for our analysis. Continuum emission from higher levels of hydrogen in the flare atmosphere, and effects from chromospheric back-warming such as discussed in \cite{2006ApJ...641.1210X}, likely provide the greatest changes to the red optical and NIR flare SED.

\begin{figure}[b]
\centering
\includegraphics[width=3.65in]{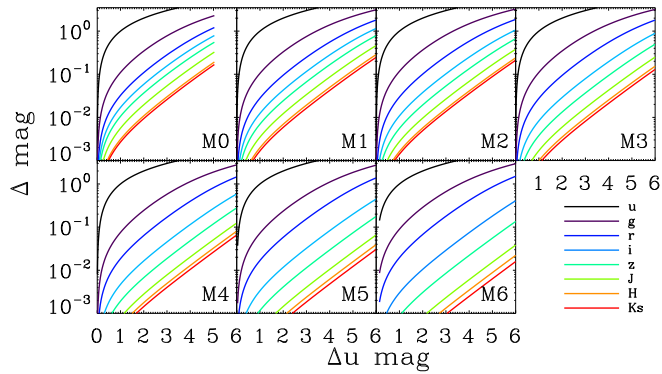}
\caption{The transformation between the predicted $u$-band response and the $grizJHK_s$-band response for each spectral type template.}
\label{dmag}
\end{figure}

To investigate the effect of the flare model, we scale it by a surface coverage fraction, ranging from 5\e{-7} to 0.5,  and add it to the M0--M6 template spectra. The photometric response to flares in each filter was then calculated by convolving the SDSS and 2MASS filter transmission curves \citep{1996AJ....111.1748F, 2003AJ....126.1090C}  to the M0--M6 model spectra with and without the flare. In Figure \ref{dmag} we show the transformations between the $\Delta u$-band and the $\Delta (grizJHK_s)$ band amplitudes for all the flare models we ran.

The top panel of Figure \ref{model} shows the changes in each photometric bandpass as computed by our model for a flare on an M3 star with 1\% surface coverage for the Balmer continuum and 0.1\% coverage for the blackbody component. The resulting increases in flux for the $ugrizJHK_s$ bands were approximately (2.7, 0.62, 0.19, 0.06, 0.03, 0.01, 0.01, 0.01) magnitudes respectively.

The extremely blue response of the flare is due to a contrast effect between the intrinsically blue flare model and the underlying red stellar spectrum. Since the M3 spectrum is much redder than the flare model, even large amplitude flares do not give a strong singal in the reddest filters. However, K09 recovered a very bright flare in the Stripe 82 sample, visible in all five passband light curves, on an active M6 star. This flare had magnitude excursions of (5.50, 3.07, 2.34, 0.77, 0.29) for the $ugriz$ bands respectively. Our M6 model, scaled to match the $u$-band excursion, predicted $ugriz$ magnitude excursions of (5.5, 2.5, 1.2, 0.3, 0.1). In the 2MASS $JHK_s$ filters we predicted excursions of  approximately (0.03, 0.01, 0.01). The coverage for this simulated flare was 0.06\% of the stellar surface. As the two-component model does not include detailed effects, as discussed above, the model excursions agree with the observations to within a factor of two in flux. The K09 flare may also have evolved in temperature over the course of the SDSS observations \citep{2008A&A...487..293F}. We allowed the blackbody temperature within our model to vary from 8,000K to 13,000K. Lower temperatures yielded only a marginally improved fit for the $r$-band, and required a larger effective surface area for the flare.

\section{Model Flare Recovery}
\begin{figure}[b]
\centering
\includegraphics[width=3.5in]{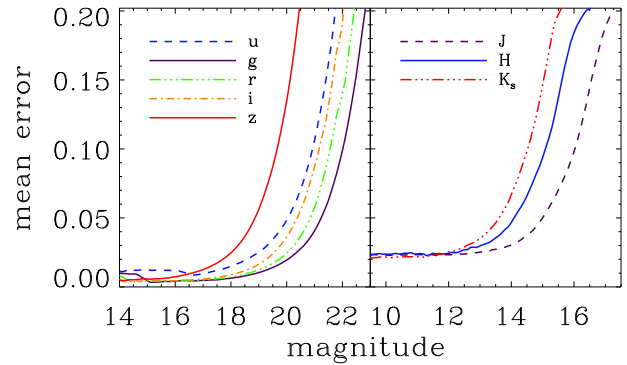}
\caption{Mean photometric error as a function of apparent magnitude for all the Stripe 82 and Cal-PSWDB measurements in our database.}
\label{meanerror}
\end{figure}

Using our flare model with the M3 spectral template, we investigated the efficiency of recovering flares from our sample with the $\Phi$ statistic. We first computed the sigma-clipped mean photometric error profiles as a function of observed magnitude for each of the $ugrizJHK_s$ bands in increments of 0.1 mag, shown in Figure \ref{meanerror}. For each of the M3 stars in our NIR time domain sample, we simulated 200 artificial flares using our model from \S5, with a logarithmically increasing surface area coverage fraction from $10^{-7}$ to $10^{-1}$. The mean error profiles were then used to select an appropriate photometric error for the resultant apparent magnitudes of each simulated flare in all eight passbands. Finally, we calculated the $\Phi$ values for each simulated flare on every star.

\begin{figure*}[]
\centering
\includegraphics[width=7in]{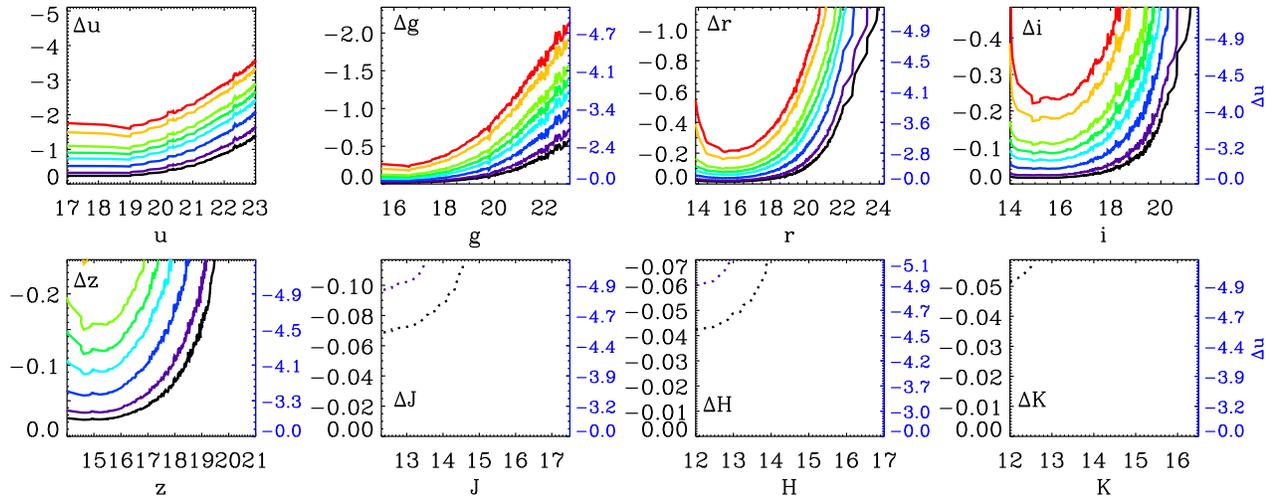}
\caption{Lines of constant $\Phi$ as a function of both mean apparent magnitude (distance) and change in magnitude (flare amplitude) in each filter, using the M3 spectral flare model above, and the mean photometric error profiles from Figure \ref{meanerror}. The predicted change in $u$-band which corresponds to the change in each other passband is shown on the right axis in blue. Contours from black to red are $\Phi$=(50, 100, 300, 700, 1200, 2000, 5000, 8000). These represent our detection thresholds for a given flare amplitude as a function of apparent magnitude in these surveys. The dotted-line threshold contours for 2MASS filters are $\Phi$=5 and 10.}
\label{response}
\end{figure*}

The result was a ``$\Phi$ profile'', which was a function of both the quiescent apparent magnitude of the star, and the flare amplitude. These represent the two competing physical quantities, the distance to the star and the size of the flare, which conspire to limit our detection of small flares or flares on distant stars. This envelope is shown in Figure \ref{response} as a function of quiescent magnitude on the x-axis, with $\Phi$ contour values increasing from black to red. The $\Phi$ profile was smoothed with a kernel of 0.1 magnitudes. Because flares are typically characterized by their enhancement of blue flux, the corresponding predicted $u$-band excursion from our model is shown on the right axis in blue for the $grizJHK_s$ bands. 

Using our described $\Phi$ thresholds from \S4, we found that small flares in the $u$-band ($\Delta u\sim 0.5$mag) would be recovered for every star in the sample with a quiescent magnitude of $u\le20.5$. The completeness is a direct function of the photometric error profile, which parallels the $\Phi$ threshold curves shown in Figure \ref{response}. Interestingly, note that in the $griz$ bands in Figure \ref{response}, the $\Phi$ profiles ``turn up'' for bright stars, since the error profiles begin to increase as stars approach the saturation limit for the SDSS. Consider a star with a moderate quiescent luminosity in the $u$-band (e.g. $u\sim17$). A very large flare such those observed in \cite{slhadleo} and \cite{k10} would result in an observed $u$-band magnitude too bright for the SDSS photometry, and would therefore be excluded from our analysis. This would be possible in the $g$, and even $r$-band, for bright stars with very large flares. Incompleteness is therefore also expected at the brightest flare luminosities, although these events are  rare with duty cycles estimated to be $\sim$1 per month on most active stars \citep{hilton10}.

The error profiles in the $g$, $r$, and $i$ filters plotted in Figure \ref{meanerror} are best suited to monitor small amplitude variations such as eclipses and spot modulation. Figures \ref{response} and \ref{dmag} clearly show that we are only completely sampling flares larger than $\Delta u\sim2$ in these bands for most of our sample, and thus the $u$-band is the most suitable to search for the frequent low-amplitude flares in these stars.

As expected from the model discussion above, only very large flares would be detectable in the NIR passbands for the brightest stars in our sample using the 2MASS photometric error profile. Our model predicts that a flare with a nearly $\sim5$ magnitude increase in the $u$-band would be required to detect any significant increase in flux in the Cal-PSWDB for a nearby M3 type star. No simulated flares on M3 stars showed enough increase in flux to register as a $\Phi=50$ response in the $JHK_s$ bands, however. The flare emission in the NIR must therefore be greatly influenced by physics not included in our two-component model, and definitive simultaneous detection of flares in the NIR and optical is needed to calibrate this model.

\subsection{Optimal NIR Flare Detection}
Our analysis has extrapolated a well-known two-component flare model into a new wavelength regime, and we therefore provide a prediction for the ``best-case'' to detect M dwarf flares in the NIR. An enhancement of $\Delta J= 0.01$ magnitudes is easily observable for many nearby M dwarfs with present day ground based differential photometry. Using the $\Delta u$ to $\Delta J$ flare amplitude transformations shown in Figure \ref{dmag}, we estimated the $\Delta u$ enhancement for each spectral type required to achieve $\Delta J = 0.01$. For spectral types M--M6 this yielded a $\Delta u=$(1.4, 1.8, 2.0, 2.5, 3.3, 4.0, 4.5) mag. The changing shape in the underlying stellar SED in the NIR indicates that $J$-band flares are detectable from smaller $\Delta u$ flares at earlier spectral types.

The frequency of flaring has been previously found to increase with spectral type, such as in our Figure \ref{ffrac}. The characteristic lifetime of magnetic activity is also highly correlated with spectral type \citep{west08}, and later spectral types M dwarfs are expected to show not only more, but larger amplitude flares (Hilton 2011 PhD Thesis). This highlights the contrasting effects at work in detecting broadband NIR flares: Persistently active M0 stars that show large amplitude flares are exceedingly rare, as compared to those at later spectral types.  The higher bolometric luminosity of earlier spectral type stars additionally means much larger energy flares are needed to achieve these large amplitudes.

\begin{figure}[]
\centering
\includegraphics[width=3.5in]{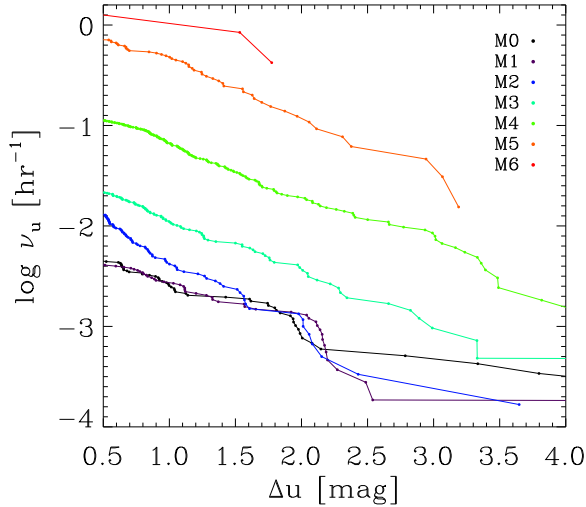}
\caption{The observed flare frequency distribution for $u$-band flares at each spectral type. Each distribution was fit with a linear polynomial in the range $0.4 \le \Delta u \le 3.0$.}
\label{ffd_sptype}
\end{figure}

\begin{figure}[]
\centering
\includegraphics[width=3.5in]{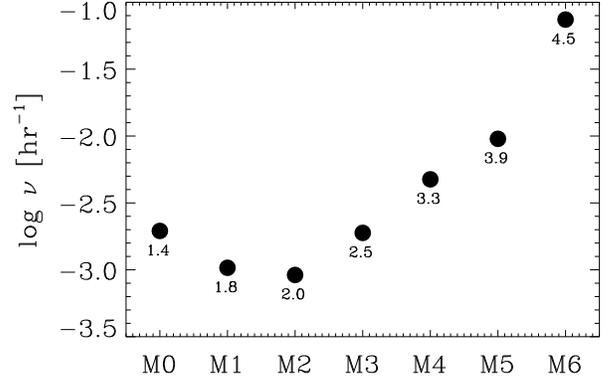}
\caption{The rate per hour at which a $\Delta J=0.01$ mag amplitude flare is expected to be seen on each spectral type, based on our model and measured flare rates. The numbers below each point are the $\Delta u$ amplitudes that our flare model predicts from a $\Delta J=0.01$ mag flare at each spectral type.}
\label{rate_sptype}
\end{figure}

Figure \ref{ffd_sptype} shows the observed $u$-band flare frequency distribution for each spectral type bin as a function of $\Delta u$ magnitude. The total rate of flaring increased with spectral type, and the slopes were very similar for all the distributions, in agreement with results from E. J. Hilton (2011 PhD Thesis). A linear fit was found for each log$_{10} \nu(\Delta u)$ distribution between $0.4\le\Delta u\le3.0$ to avoid false-positives at the low-amplitude end, and incompleteness at high $\Delta u$ amplitudes. The predicted frequency at which we expect to observe a $\Delta J=0.01$ mag flare, from the corresponding $\Delta u$ amplitudes, at each spectral type was then calculated from these fits, and is shown in Figure \ref{rate_sptype}. The optimal spectral type to search for these NIR flare detections is in our latest type stars, M6. The slight up-turn at M0 is due to the more favorable color-contrast between the flare and the star. The overall trend from M1--M6, however, shows that the strongest effect for all spectral types is the total flaring duty cycle. This provides one possible reason that the recovered flare rates are higher than expected for M0--M1 stars in Figure \ref{ffrac}. A new method for calculating the duty cycle of a given amplitude excursion, known as the Magnitude Likelihood Distribution, has been recently put forward by E. J. Hilton (2011 PhD Thesis). Using these Magnitude Likelihood Distributions we predict a peak duty cycle of $\sim3\e{-4}$ for the magnetically active M4 and M5 stars. For the 54 sec exposures of SDSS, this is equal to a frequency of $\nu\approx3\e{-2}$ hr$^{-1}$, which is consistent with our estimation. 
We would encourage future NIR flare observers to monitor bright/nearby targets in the latest spectral types possible, at least M4 or later. Promising stars might include GJ 1061, CN Leo (Wolf 359), and especially the highly active YZ CMi.

\section{Summary and Conclusions}
We have shown that the signatures of flares from M dwarfs are not confined to the bluest bandpasses in large scale photometric surveys. There is significant response from medium to large flares in $gri$ bands, and the largest flares are observable in $z$ and in some cases into the NIR passbands. This is the first statistical characterization of M dwarf flares in the NIR regime.

A self-consistent result between all eight filters shows that while the frequency of flares decreases with  increasing luminosity in all filters, the total number of flares recovered decreases with wavelength. This suggests that like our simplified flare model the underlying SED for flares of all sizes is intrinsically blue. Furthermore, the total rate of flares in all optical bands appears to increase with increasing spectral type, but low-number statistics prevent us from robustly exploring this distribution at NIR wavelengths. The slope of the flare frequency distribution is found to be insensitive to the flare detection threshold below $\sim10^{29}$ erg s$^{-1}$.

We have also presented an implementation of the two-component spectral flare model from \cite{k10} to make predictions of the red-optical and NIR response of flares in M dwarfs. These conversions will be useful in future broadband photometric studies of M dwarfs to predict the contamination from flares across a wider wavelength space. Flares in the $J$-band are predicted to be detected most frequently in spectral types from M4 to M6, with rates of $\sim2\e{-2}$ hr$^{-1}$.

Medium sized flares larger than 0.5 magnitudes in $u$ were found to occur at a rate of one per few thousand SDSS exposures for M dwarfs. Low-amplitude flare rates are limited by the photometric precision of the survey in question, and LSST will detect even smaller amplitude flares from these stars, and will increase the rate at which variability from these sources is detected.  LSST will also employ a $y$-filter, with a central wavelength between that of $z$ and $J$. Using the ``$y4$'' filter curve for LSST, our two-component flare model from Figure \ref{model} predicts a flare amplitude of $\Delta y4\sim 0.02$ mag. The increased photometric sensitivity will therefore allow LSST to routinely detect large flares in this NIR bandpass for bright stars, providing useful constraints on the SED structure of flares at long wavelengths

Finally, we have shown that the intrinsic SED of a flare, convolved with the photometric characteristics of a given survey, can greatly affect the sample of flares recovered. The cadence and integration times will also affect the recovery of flares. Short exposures such as in 2MASS capture the near instantaneous luminosity of an event, while longer exposure times may encapsulate entire flare events. 
We anticipate a day when this problem can be inverted, using orders of magnitude larger samples of flare candidate epochs in many bands, as well as spectra, from next-generation surveys to probe the underlying flare SED (or its deviations) across the population of M dwarfs in the Galaxy.

\acknowledgements 
The authors would like to thank C. MacLeod and J. Ruan for their insight in characterizing astrophysical variability, B. Tofflemire and J. Wisniewski for fruitful discussions regarding NIR flares, and P. Kundurthy for his assistance in converting times to BJD. ACB and JRAD acknowledge support from NASA ADP grant NNX09AC77G. EJH, SLH, and AFK acknowledge support from NSF grant AST 08-07205.

This publication makes use of data products from the Two Micron All Sky Survey, which is a joint project of the University of Massachusetts and the Infrared Processing and Analysis Center, California Institute of Technology, funded by the National Aeronautics and Space Administration and the National Science Foundation.

This research has made use of the SIMBAD database, operated at CDS, Strasbourg, France.

Funding for the SDSS and SDSS-II has been provided by the Alfred P. Sloan Foundation, the Participating Institutions, the National Science Foundation, the U.S. Department of Energy, the National Aeronautics and Space Administration, the Japanese Monbukagakusho, the Max Planck Society, and the Higher Education Funding Council for England. The SDSS Web Site is \url{http://www.sdss.org/}.

The SDSS is managed by the Astrophysical Research Consortium for the Participating Institutions. The Participating Institutions are the American Museum of Natural History, Astrophysical Institute Potsdam, University of Basel, University of Cambridge, Case Western Reserve University, University of Chicago, Drexel University, Fermilab, the Institute for Advanced Study, the Japan Participation Group, Johns Hopkins University, the Joint Institute for Nuclear Astrophysics, the Kavli Institute for Particle Astrophysics and Cosmology, the Korean Scientist Group, the Chinese Academy of Sciences (LAMOST), Los Alamos National Laboratory, the Max-Planck-Institute for Astronomy (MPIA), the Max- Planck-Institute for Astrophysics (MPA), New Mexico State University, Ohio State University, University of Pittsburgh, University of Portsmouth, Princeton University, the United States Naval Observatory, and the University of Washington.


\end{document}